\documentclass[10pt,a4paper,onecolumn]{article}
\usepackage{marginnote}
\usepackage{graphicx}
\usepackage[rgb,dvipsnames]{xcolor}
\usepackage{authblk,etoolbox}
\usepackage{titlesec}
\usepackage{calc}
\usepackage{tikz}
\usepackage[pdfa]{hyperref}
\usepackage{hyperxmp}
\hypersetup{%
    unicode=true,
    pdfapart=3,
    pdfaconformance=B,
    pdftitle={Reggae: A Parametric Tuner for PBJam, and a Visualization
Tool for Red Giant Oscillation Spectra},
    pdfauthor={J. M. Joel Ong, Martin B. Nielsen, Emily J. Hatt, Guy R.
Davies},
    pdfpublication={Journal of Open Source Software},
    pdfpublisher={Open Journals},
    pdfissn={2475-9066},
    pdfpubtype={journal},
    pdfvolumenum={},
    pdfissuenum={},
    pdfdoi={10.21105/joss.06588},
    pdfcopyright={Copyright (c) 2024, J. M. Joel Ong, Martin B. Nielsen,
Emily J. Hatt, Guy R. Davies},
    pdflicenseurl={http://creativecommons.org/licenses/by/4.0/},
    colorlinks=true,
    linkcolor=[rgb]{0.0, 0.5, 1.0},
    citecolor=Blue,
    urlcolor=[rgb]{0.0, 0.5, 1.0},
    breaklinks=true
}
\immediate\pdfobj stream attr{/N 3} file{sRGB.icc}
\pdfcatalog{%
  /OutputIntents [
    <<
      /Type /OutputIntent
      /S /GTS_PDFA1
      /DestOutputProfile \the\pdflastobj\space 0 R
      /OutputConditionIdentifier (sRGB)
      /Info (sRGB)
    >>
  ]
}

\usepackage{caption}
\usepackage{orcidlink}
\usepackage{tcolorbox}
\usepackage{amssymb,amsmath}
\usepackage{ifxetex,ifluatex}
\usepackage{seqsplit}
\usepackage{xstring}

\usepackage{float}
\let\origfigure\figure
\let\endorigfigure\endfigure
\renewenvironment{figure}[1][2] {
    \expandafter\origfigure\expandafter[H]
} {
    \endorigfigure
}

\usepackage{fixltx2e} 

\NewDocumentCommand\citeproctext{}{}
\NewDocumentCommand\citeproc{mm}{%
  \begingroup\def\citeproctext{#2}\cite{#1}\endgroup}
\makeatletter
 \let\@cite@ofmt\@firstofone
 \def\@biblabel#1{}
 \def\@cite#1#2{{#1\if@tempswa , #2\fi}}
\makeatother
\newlength{\cslhangindent}
\newlength{\csllabelwidth}
\newenvironment{CSLReferences}[2] 
 {\begin{list}{}{%
  \setlength{\itemindent}{0pt}
  \setlength{\leftmargin}{0pt}
  \setlength{\parsep}{0pt}
  \ifodd #1
   \setlength{\leftmargin}{\cslhangindent}
   \setlength{\itemindent}{-1\cslhangindent}
  \fi
  \setlength{\itemsep}{#2\baselineskip}}}
 {\end{list}}
\usepackage{calc}

\usepackage[top=3.5cm, bottom=3cm, right=1.5cm, left=1.0cm,
            headheight=2.2cm, reversemp, includemp, marginparwidth=4.5cm]{geometry}

\titleformat{\section}
  {\normalfont\sffamily\Large\bfseries}
  {}{0pt}{}
\titleformat{\subsection}
  {\normalfont\sffamily\large\bfseries}
  {}{0pt}{}
\titleformat{\subsubsection}
  {\normalfont\sffamily\bfseries}
  {}{0pt}{}
\titleformat*{\paragraph}
  {\sffamily\normalsize}

\usepackage{fancyhdr}
\makeatletter
\let\ps@plain\ps@fancy
\fancyheadoffset[L]{4.5cm}
\fancyfootoffset[L]{4.5cm}

\definecolor{linky}{rgb}{0.0, 0.5, 1.0}

\newtcolorbox{repobox}
   {colback=red, colframe=red!75!black,
     boxrule=0.5pt, arc=2pt, left=6pt, right=6pt, top=3pt, bottom=3pt}

\newcommand{\ExternalLink}{%
   \tikz[x=1.2ex, y=1.2ex, baseline=-0.05ex]{%
       \begin{scope}[x=1ex, y=1ex]
           \clip (-0.1,-0.1)
               --++ (-0, 1.2)
               --++ (0.6, 0)
               --++ (0, -0.6)
               --++ (0.6, 0)
               --++ (0, -1);
           \path[draw,
               line width = 0.5,
               rounded corners=0.5]
               (0,0) rectangle (1,1);
       \end{scope}
       \path[draw, line width = 0.5] (0.5, 0.5)
           -- (1, 1);
       \path[draw, line width = 0.5] (0.6, 1)
           -- (1, 1) -- (1, 0.6);
       }
   }

\patchcmd{\@maketitle}{center}{flushleft}{}{}
\patchcmd{\@maketitle}{center}{flushleft}{}{}
\patchcmd{\@maketitle}{\LARGE}{\LARGE\sffamily}{}{}
\def\maketitle{{%
  
  \AB@maketitle}}
\renewcommand\AB@affilsepx{ \protect\Affilfont}
\renewcommand\AB@affilnote[1]{{\bfseries #1}\hspace{3pt}}
\renewcommand{\affil}[2][]%
   {\newaffiltrue\let\AB@blk@and\AB@pand
      \if\relax#1\relax\def\AB@note{\AB@thenote}\else\def\AB@note{#1}%
        \setcounter{Maxaffil}{0}\fi
        \begingroup
        \let\href=\href@Orig
        \let\protect\@unexpandable@protect
        \def\thanks{\protect\thanks}\def\footnote{\protect\footnote}%
        \@temptokena=\expandafter{\AB@authors}%
        {\def\\{\protect\\\protect\Affilfont}\xdef\AB@temp{#2}}%
         \xdef\AB@authors{\the\@temptokena\AB@las\AB@au@str
         \protect\\[\affilsep]\protect\Affilfont\AB@temp}%
         \gdef\AB@las{}\gdef\AB@au@str{}%
        {\def\\{, \ignorespaces}\xdef\AB@temp{#2}}%
        \@temptokena=\expandafter{\AB@affillist}%
        \xdef\AB@affillist{\the\@temptokena \AB@affilsep
          \AB@affilnote{\AB@note}\protect\Affilfont\AB@temp}%
      \endgroup
       \let\AB@affilsep\AB@affilsepx
}
\makeatother

\renewcommand\Affilfont{\sffamily\small\mdseries}
\ifnum 0\ifxetex 1\fi\ifluatex 1\fi=0 
  \usepackage[T1]{fontenc}
  \usepackage[utf8]{inputenc}

\else 
  \ifxetex
    \usepackage{mathspec}
    \usepackage{fontspec}

  \else
    \usepackage{fontspec}
  \fi
  \defaultfontfeatures{Scale=MatchLowercase}
  \defaultfontfeatures[\sffamily]{Ligatures=TeX}
  \defaultfontfeatures[\rmfamily]{Ligatures=TeX,Scale=1}

\fi
\IfFileExists{upquote.sty}{\usepackage{upquote}}{}
\IfFileExists{microtype.sty}{%
\usepackage{microtype}
\UseMicrotypeSet[protrusion]{basicmath} 
}{}

\PassOptionsToPackage{usenames,dvipsnames}{color} 
\ifLuaTeX
\usepackage[bidi=basic]{babel}
\else
\usepackage[bidi=default]{babel}
\fi
\babelprovide[main,import]{american}

\def\languageshorthands#1{}

\usepackage{graphicx,grffile}
\makeatletter
\def\maxwidth{\ifdim\Gin@nat@width>\linewidth\linewidth\else\Gin@nat@width\fi}
\def\maxheight{\ifdim\Gin@nat@height>\textheight\textheight\else\Gin@nat@height\fi}
\makeatother
\setkeys{Gin}{width=\maxwidth,height=\maxheight,keepaspectratio}
\IfFileExists{parskip.sty}{%
\usepackage{parskip}
}{
\setlength{\parindent}{0pt}
\setlength{\parskip}{6pt plus 2pt minus 1pt}
}
\ifx\paragraph\undefined\else
\let\oldparagraph\paragraph
\renewcommand{\paragraph}[1]{\oldparagraph{#1}\mbox{}}
\fi
\ifx\subparagraph\undefined\else
\let\oldsubparagraph\subparagraph
\renewcommand{\subparagraph}[1]{\oldsubparagraph{#1}\mbox{}}
\fi
\ifLuaTeX
  \usepackage{selnolig}  
\fi

\title{Reggae: A Parametric Tuner for PBJam, and a Visualization Tool
for Red Giant Oscillation Spectra}

\author[1,2%
\ensuremath\mathparagraph]{J. M. Joel Ong%
  \,\orcidlink{0000-0001-7664-648X}\,%
}
\author[3%
]{Martin B. Nielsen%
  \,\orcidlink{0000-0001-9169-2599}\,%
}
\author[3%
]{Emily J. Hatt%
  \,\orcidlink{0000-0002-1389-1549}\,%
}
\author[3%
]{Guy R. Davies%
  \,\orcidlink{0000-0002-4290-7351}\,%
}

\affil[1]{NASA Hubble Fellow}
\affil[2]{Institute for Astronomy, University of Hawai`i, 2680 Woodlawn
Drive, Honolulu, HI 96822, USA}
\affil[3]{School of Physics and Astronomy, University of Birmingham,
Birmingham B15 2TT, UK}
\affil[$\mathparagraph$]{Corresponding author}
\date{\vspace{-2.5ex}}

\begin{document}
\maketitle

\marginpar{

  \begin{flushleft}
  \sffamily\small

  {\bfseries DOI:} \href{https://doi.org/10.21105/joss.06588}{\color{linky}{10.21105/joss.06588}}

  \vspace{2mm}
    {\bfseries Software}
  \begin{itemize}
    \setlength\itemsep{0em}
    \item \href{https://joss.theoj.org/papers/e6adb7a3b7cabe398f6c23297da1d3b3}{\color{linky}{Review}} \ExternalLink
    \item \href{https://github.com/darthoctopus/reggae}{\color{linky}{Repository}} \ExternalLink
    \item \href{https://doi.org/10.5281/zenodo.12730547}{\color{linky}{Archive}} \ExternalLink
  \end{itemize}

  \vspace{2mm}
  
    \par\noindent\hrulefill\par

  \vspace{2mm}

  {\bfseries Editor:} \href{https://github.com/dfm}{Dan
Foreman-Mackey} \ExternalLink
  \,\orcidlink{0000-0002-9328-5652} \\
  \vspace{1mm}
    {\bfseries Reviewers:}
  \begin{itemize}
  \setlength\itemsep{0em}
    \item \href{https://github.com/sblunt}{@sblunt}
    \item \href{https://github.com/sybreton}{@sybreton}
    \end{itemize}
    \vspace{2mm}
  
    {\bfseries Submitted:} 07 February 2024\\
    {\bfseries Published:} 13 July 2024

  \vspace{2mm}
  {\bfseries License}\\
  Authors of papers retain copyright and release the work under a Creative Commons Attribution 4.0 International License (\href{https://creativecommons.org/licenses/by/4.0/}{\color{linky}{CC BY 4.0}}).

  \end{flushleft}
}

\section{Summary}\label{summary}

\texttt{PBjam} (\citeproc{ref-2021AJ....161...62N}{Nielsen et al.,
2021}) is a software instrument for fitting normal modes
(``peakbagging'') in power spectra from space-based photometry of
solar-like oscillators (e.g.
\citeproc{ref-garcia_asteroseismology_2019}{Garc\'ia \& Ballot, 2019}).
Its upcoming second release (\citeproc{ref-pbjam2}{Nielsen et al., in
prep.}) supplements the simple power-spectrum model used in the first
version --- which included only radial and quadrupole (\(\ell = 0, 2\))
modes --- to additionally constrain other features (e.g.
\citeproc{ref-2023Aux26A...676A.117N}{Nielsen et al., 2023}). Dipole
(\(\ell = 1\)) modes, which had been specifically excluded in the
initial version of the tool owing to their complexity, are now
specifically included. Since the primary samples of the PLATO mission
consist mainly of main-sequence and subgiant stars
(\citeproc{ref-plato}{Rauer et al., 2024}), \texttt{PBjam} implements a
single parameterisation of dipole mixed-mode frequencies --- as
described by their overtone spacings, boundary conditions, and other
stellar properties --- that reduces to pure p-modes in the former, and
is suitable to the latter, outside the red-giant ``asymptotic'' regime.
In keeping with the overall philosophy of \texttt{PBjam}'s design for
\(\ell = 0,2\), \texttt{PBjam} 2 will specify prior distributions on
these parameters empirically, through predetermined values found for
existing samples of solar-like oscillators. While the red-giant
asymptotic regime has been extensively characterised observationally,
the nonasymptotic construction for subgiants here has not, requiring us
to construct this prior sample ourselves. To assist in this task, we
built a tool --- \texttt{Reggae}--- to manually fine-tune and fit the
dipole-mode model, and check the quality of both our initial guesses and
fitted solutions.

\section{Statement of Need}\label{statement-of-need}

Before mode frequencies may be extracted from the power spectrum,
specific peaks in it must be identified as dipole modes. An important
part of this identification is visual assessment of how well the
predicted mode frequencies correspond to actually observed peaks.
\texttt{Reggae} produces these visualisations from user-supplied trial
values. This is useful for checking solutions of, e.g., the period
spacing \(\Delta\Pi_1\) --- inaccurate values result in slanted ridges
on period-echelle diagrams (e.g.
\citeproc{ref-mosser_probing_2012}{Mosser et al., 2012}), much like with
inaccurate \(\Delta\nu\) in traditional frequency-échelle diagrams (e.g.
\citeproc{ref-aerts_asteroseismology_2010}{Aerts et al., 2010}).
Similarly, rotational splittings become easily identifiable (e.g.
\citeproc{ref-gehan_core_2018}{Gehan et al., 2018}), as are any
perturbations due to magnetic fields (\citeproc{ref-hatt}{Hatt et al.,
submitted to MNRAS}).

Since these global parameters must be supplied for dipole-mode
identification, we have constrained them for a preliminary sample of
subgiants (\citeproc{ref-pbjam2}{Nielsen et al., in prep.}), and also
for a large sample of low-luminosity red giants
(\citeproc{ref-hatt}{Hatt et al., submitted to MNRAS}). We found
\texttt{Reggae} very helpful both for these tuning and visualisation
tasks, and also as a didactic aid to understanding the dipole mixed-mode
parameters. Moreover, no other tools currently exist for performing
these tasks in the nonasymptotic parameterisation that \texttt{PBjam}
will use. As such, we release \texttt{Reggae} publicly in advance of the
second \texttt{PBjam} version, as we believe the community will benefit
from access to such a visualisation tool. This will also assist future
users of \texttt{PBjam} in devising ad-hoc prior constraints on the
mixed-mode parameters, should they wish to perform mode identification
for anomalous stars.

\section{Modeling the Oscillation
Spectrum}\label{modeling-the-oscillation-spectrum}

\texttt{Reggae} uses individually-fitted modes from \texttt{PBjam} to
construct a model of the \(\ell=2,0\) modes, which is then divided out
of the signal-to-noise spectrum; this allows the optimization and
visualization of the \(\ell=1\) mode identification to be performed
independently, and far more simply. The dipole p-mode frequencies are
parameterised identically to \texttt{PBjam}, with a small frequency
offset \(d_{01} \times \Delta\nu\) to account for imperfections in this
idealised asymptotic description.

To produce mixed modes, we must specify both pure g-mode frequencies ---
which we describe using a period spacing \(\Delta\Pi_1\), a g-mode phase
offset \(\epsilon_g\), and an analogous curvature parameter \(\alpha_g\)
to that used in the p-mode parameterisation --- as well as coupling
between the p- and g-modes. For this \texttt{PBJam} will adopt the
matrix-eigenvalue parameterisation of Deheuvels \& Michel
(\citeproc{ref-deheuvels_insights_2010}{2010}), supplemented with a
secondary inner-product matrix as described in Ong \& Basu
(\citeproc{ref-ong_semianalytic_2020}{2020}) to account for the
nonorthogonality of the notional pure p- and g-mode eigenfunctions. This
parameterisation is used instead of the classical asymptotic description
(e.g. \citeproc{ref-1979PASJ...31...87S}{Shibahashi, 1979}) in light of
its intended application to subgiants specifically. Numerically, these
matrices are scaled from values supplied by a reference MESA model (from
the grid of \citeproc{ref-lindsay}{Lindsay et al., 2024}) using
parameters \(p_\mathrm{L}\) and \(p_\mathrm{D}\). The correspondence
between these matrices and the classical coupling strength \(q\) is
described in Ong \& Gehan (\citeproc{ref-ong_rotation_2023}{2023}).
Rotation in the p- and g-mode cavities are separately parameterised with
\(\log \Omega_\mathrm{p}\) and \(\log \Omega_\mathrm{g}\), and a shared
inclination parameter \(i\), with rotating mixed modes computed fully
accounting for near-degeneracy effects (e.g.
\citeproc{ref-ong_rotation_2022}{Ong et al., 2022}).

\texttt{Reggae} fine-tunes these parameters by numerical optimization,
which requires a model of the power spectral density (PSD) that can be
compared to the observed residual spectrum. This model is a sum of
Lorentzian profiles, one for each of the predicted dipole modes. Their
linewidths are artificially broadened to a fraction of \(\Delta\nu\),
smoothing over local minima in the likelihood function. Their heights
follow the same Gaussian envelope as \texttt{PBjam}'s model for the
\(\ell=2,0\) pairs, with additional modulation by mixing fractions
\(\zeta\) from mode coupling.

\begin{figure}
\centering
\includegraphics{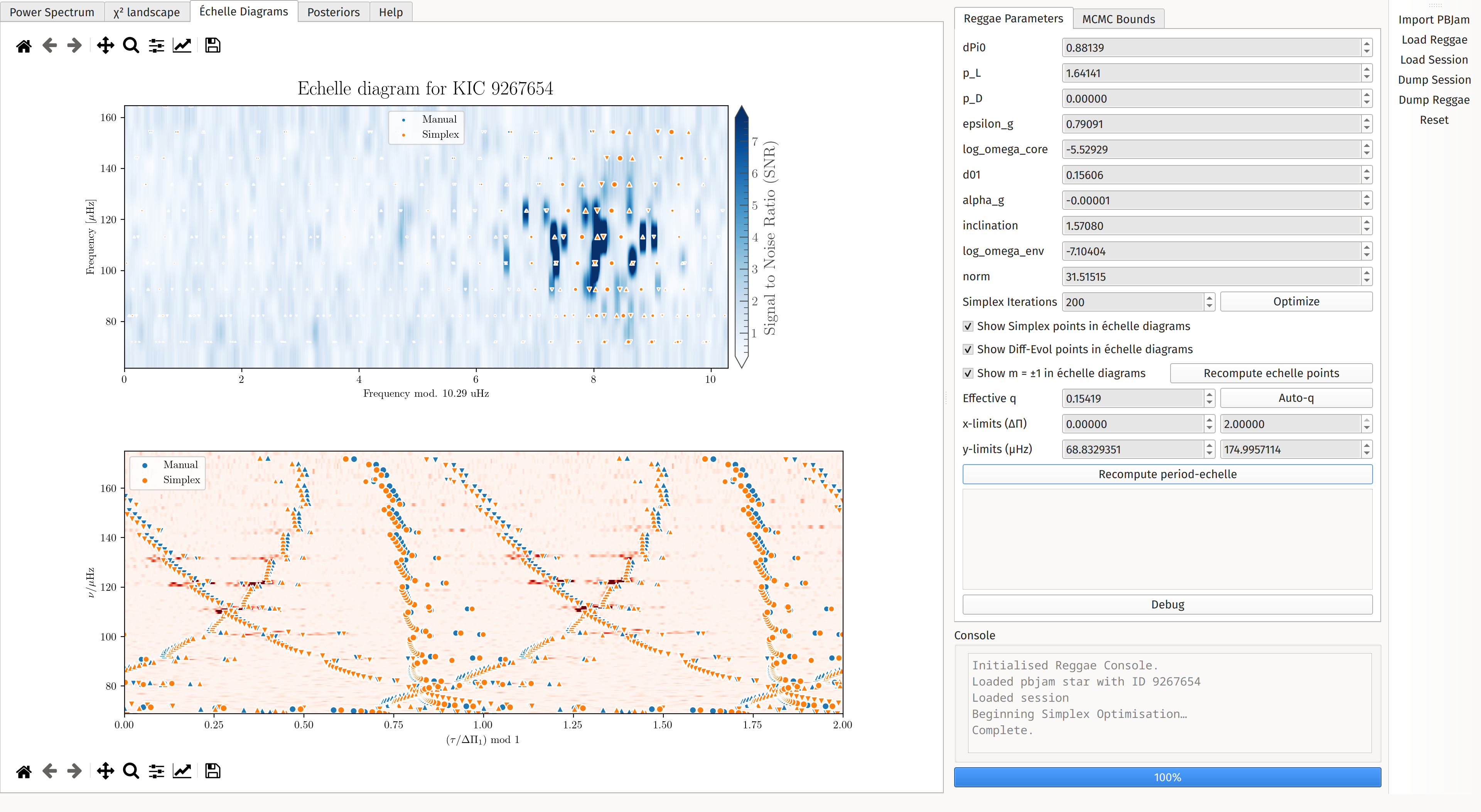}
\caption{Screenshot of the GUI showing visualisation panel and manual
inputs.\label{fig:screenshot}}
\end{figure}

These visualization and tuning features are operated through a graphical
user interface (GUI), illustrated in \autoref{fig:screenshot}. The
visualisation tools are provided on the left of the interface. Manual
guesses and parameter bounds provide initial guesses for simplex or
genetic-algorithm optimization. Alternatively all parameters can be
sampled at once using the Dynesty nested sampling package
(\citeproc{ref-dynesty}{Koposov et al., 2022}).

\section{Acknowledgments}\label{acknowledgments}

JMJO acknowledges support from NASA through the NASA Hubble Fellowship
grants HST-HF2-51517.001-A, awarded by STScI, which is operated by the
Association of Universities for Research in Astronomy, Incorporated,
under NASA contract NAS5-26555. MBN acknowledges support from the UK
Space Agency.

\section*{References}\label{references}
\addcontentsline{toc}{section}{References}

\phantomsection\label{refs}

\end{document}